\documentclass[aip,apl]{revtex4}
\usepackage{graphicx}

\begin{document}
\author{Gerardo Algara-Siller}
\altaffiliation{Contributed equally to this work}
\author{Simon Kurasch}
\altaffiliation{Contributed equally to this work}
\author{Mona Sedighi}
\author{Ossi Lehtinen}
\author{Ute Kaiser}
\email{ute.kaiser@uni-ulm.de}
\affiliation{Central Facility for Electron Microscopy, Group of Electron Microscopy of Materials Science, Ulm University}

\title{The pristine atomic structure of MoS$_2$ monolayer protected from electron radiation damage by graphene}

\begin{abstract}
Materials can, in principle, be imaged at the level of individual atoms with aberration corrected transmission electron microscopy. However, such resolution can be attained only with very high electron doses. Consequently, radiation damage is often the limiting factor when characterizing sensitive materials. Here, we demonstrate a simple and effective method to increase the electron radiation tolerance of materials by using graphene as protective coating. This leads to an improvement of three orders of magnitude in the radiation tolerance of monolayer MoS$_2$. Further on, we construct samples in different heterostructure configurations to separate the contributions of different radiation damage mechanisms.
\end{abstract}


\maketitle

Radiation damage is a prevalent challenge when materials are introduced in radiation-hostile environments, either on purpose or inadvertently, {\em e.g.}, in nuclear reactors, space applications, or scientific instruments such as in transmission electron microscopes (TEM). Nowadays in state of the art TEMs, the sample is imaged using a beam of energetic electrons with energies ranging from 20~keV~\cite{kaiser2011,sawada2010,krivanek2010} to 300~keV. Hardware aberration-correction in a TEM~\cite{haider1998,krivanek1999,kabius2009} allows -- in principle -- direct {\em in-situ} observation of single atoms~\cite{meyer2008} and single atomic columns~\cite{jia2003} of materials. However, the instrumental resolution can only be reached when the material can withstand an unlimited electron dose. In practice, this is rarely the case. Therefore the resolution in the high-resolution TEM image is nowadays predominantly limited by the electron dose the specimen can tolerate before it is damaged~\cite{rose2009}.

On the other hand, simultaneous production and observation of radiation-induced transformations in the sample allow detailed studies of radiation damage in a TEM. Traditionally second-order phenomena, such as fading of diffraction spots connected to the loss of sample crystallinity, mass-loss, changes in the energy-loss spectra, and electron beam induced x-ray yield have been used for characterization of radiation damage~\cite{reimer1984} in classical three-dimensional specimens. Two-dimensional (2D) materials, such as graphene, {\em h}-BN and MoS$_2$, however, have allowed resolving the electron-beam-induced changes atom-by-atom, making a direct observation of radiation damage at the level of the basic building blocks of matter -- the atoms -- possible~\cite{banhart2011,kotakoski2011,meyer2012,komsa2012}. A notable difference to bulk materials is that in such 2D materials surface effects are dominating, as in essence, surface is all the materials have.

Radiation damage in the TEM is divided into two different categories, where the first one is related to displacements of atoms through direct collisions between the energetic electrons and target atoms (so called knock-on damage), and the second to excitations of the electronic system of the target~\cite{egerton2004}, which can lead, {\em e.g.}, to radiolysis. Moreover, energy deposition from the electron beam can lead to heating of the specimen, and to ejection of secondary electrons. The latter may result in significant electrostatic charging in insulating materials, which can even lead to a so-called Coulomb explosion~\cite{wei2013}, where heavily charged material disintegrates due to internal electrostatic repulsion. Contaminants on the sample surface and residual gases in the vacuum of the microscope can be broken down, thus creating free radicals, which can lead to chemical etching of the sample surface~\cite{reimerkohl2008}. Interplay of damage mechanisms is possible, {\em e.g.}, if the response of the target to knock-on collisions is different in an electronically excited or charged state. Theoretical predictions of radiation damage in the TEM have been limited mostly to the simplest case of the knock-on process~\cite{kotakoski2011,meyer2012,komsa2012} since calculations on the other processes are considerably more challenging.

Many approaches have been proposed and employed for combating the adverse effects of electron radiation. On the instrumentation side, lowering the electron energy reduces knock-on damage, as there is a material specific threshold in momentum which needs to be transferred to the target atom in order to displace it from its lattice position. The cost of a lower electron energy, however, is loss in attainable resolution, and thus the aim is to work below, but close to the damage threshold energy. On the other hand, increasing the electron energy lowers the inelastic scattering cross-section, thus reducing damage related to the excitations of the electronic system of the target. Depending on which mechanism is predominant in a sample, an optimal electron energy needs to be selected~\cite{egerton2004,kaiser2011}.

An alternative route for combating radiation damage is treatment of the sample itself. Protecting a specimen by coating it with conducting and/or radiation resistant material is a well established method~\cite{reimerkohl2008}. Such coating can serve as a source of electrons which can compensate the charging effect in the sample, serve as a thermal conductor, and/or protect the sample from chemical etching~\cite{yuk2012}. Further on, when the exit surface of the sample is coated, sputtering of target material can be inhibited. As an extreme demonstration of the benefits of coating a sample, mosquito larvae were recently demonstrated to come out alive and even hatch after investigation in the vacuum of a scanning electron microscope when a coating treatment was applied~\cite{takaku2013}.

A typical drawback of coating a specimen, however, is that it obscures the specimen itself, degrades the signal to noise ratio, and consequently reduces the attainable specimen resolution. Therefore an optimal coating material would have excellent electrical and thermal conductivity, be resistant and highly transparent to electrons at the used energy, be chemically inert, and be crystalline. The last is important as the known image contribution of the crystalline lattice can be removed from the micrographs in digital post-processing, {\em e.g.} by Fourier filtering~\cite{lee2009,westenfelder2011}. A material meeting all these criteria is graphene, and it has indeed been suggested and employed as the ultimate sample substrate for TEM~\cite{longchamp2012,westenfelder2011b,pantelic2011,kaiser2011,nair2010,pantelic2012,yuk2012,lee2011,lee2009,warner2010,meyer2008}. The usefulness of graphene in reducing ion radiation damage has been reported in other studies, where significant reduction in the sputtering yield of platinum covered with graphene under ion irradiation has been predicted~\cite{standop2013}, and reduced damage in graphene sandwiched between SiO$_2$ and another graphene layer under ion bombardment has been observed~\cite{kalbac2013}.

\begin{figure}
	\begin{center}
	\includegraphics[width=.98\textwidth]{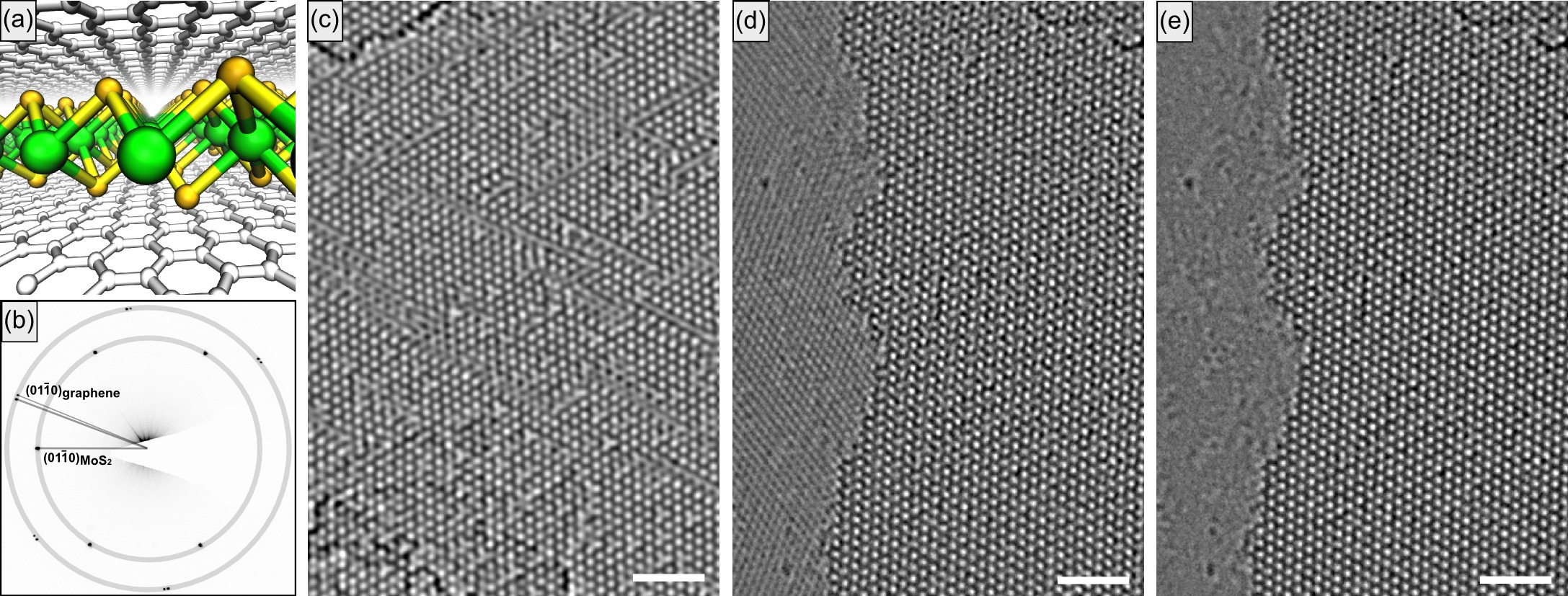}
	\caption{{\bf The effect of sandwiching MoS$_2$ single layer in between graphene layers (the G/MoS$_2$/G configuration)}. {\bf (a)}, A schematic representation of the sandwich structure, where a MoS$_2$ monolayer is confined in between two graphene layers (green spheres are Mo, and yellow spheres S, and white spheres C). {\bf (b)}, Electron diffraction pattern of the G/MoS$_2$/G sample, showing the monolayer MoS$_2$ peaks at 3.7~nm$^{-1}$ and the two graphene peaks at 4.7~nm$^{-1}$in different orientations, originating from the top and bottom layers. The ratio of the alternating intensities of the first-order reflections of the MoS$_2$ maxima is 1.20, confirming its monolayer nature~\cite{brivio2011}. {\bf (c)}, An 80~keV AC-HRTEM image of free-standing MoS$_2$ monolayer after an electron dose of $2.8\times10^8$~electrons/nm$^2$. Serious damaging of the sample can be observed. {\bf (d)}, An 80~keV AC-HRTEM image of the G/MoS$_2$/G sample after an electron dose of $2.8\times10^8$~electrons/nm$^2$. {\bf (e)}, The frame of panel {\bf (d)} after Fourier filtering the graphene contribution out. Here the electron radiation damage is dramatically reduced and, {\em e.g.}, determination of the MoS$_2$ edge structure is possible. The scale bars are 2~nm.}
	\label{firstimage}
	\end{center}
\end{figure}

In this study, we show how coating an electron beam sensitive material --- single layer MoS$_2$ --- with graphene dramatically improves its electron radiation resistance. We analyse the evolution of the electron radiation damage at atomic resolution, employing 80~kV AC-HRTEM. We show that the electron radiation resistance of the MoS$_2$ single layer is increased by nearly three orders of magnitude when it is sandwiched between two graphene layers, as compared to the free standing MoS$_2$ layer (see \ref{firstimage}). This demonstrates the effectiveness of our graphene-sandwich-approach in fighting radiation damage. In addition, by preparing samples in different heterostructure configurations we are able to separate the role of the different damage mechanisms in the material at the level of single atoms.

For this purpose we constructed samples in four different configurations, using the following procedure: A graphene flake (1-3 layers) was mechanically exfoliated from graphite, deposited on SiO$_2$(90nm)/Si, and transferred to Quantifoil grids using KOH\cite{meyer2008b}. Single layer MoS$_2$ was also cleaved from a natural crystal, and deposited on SiO$_2$/Si. In order to transfer the MoS$_2$ flake to the already transferred graphene we positioned the graphene/Quantifoil on top of the single layer MoS$_2$. Once positioned the MoS$_2$ layer was transferred via the same method as used for graphene. The process was repeated for a another graphene layer. As a result we got a sandwich structure where the MoS$_2$ layer lays in between graphene layers. The HRTEM investigations were performed in an aberration-corrected FEI TITAN 80-300 operated at 80 kV with C$_s$= 0.03 mm under Scherzer conditions. The vacuum level in the microscope was $< 10^{-7}$~mbar.

As the baseline sample, we used a free-standing MoS$_2$, which was also the subject of an earlier study~\cite{komsa2012}. For evaluating the ultimate protective power of the graphene layers, a sandwich structure was constructed (named the G/MoS$_2$/G configuration), where the MoS$_2$ layer was enclosed between graphene layers via subsequent transfer of the mechanically exfoliated layers on to a TEM grid at the same position (see \ref{firstimage}a for a schematic representation). 
For separating the effects of the different damage mechanisms, two heterostructures with a MoS$_2$ monolayer and graphene were constructed, with graphene either on top or bottom of the MoS$_2$ layer (named the G/MoS$_2$ and MoS$_2$/G configurations).

Based on light-micrographs produced during sample preparation the correct sample area was located at low magnification in the TEM in order to minimize the electron dose before characterization of the effects of the electron beam on the material. Electron diffraction patterns were recorded for verifying the monolayer nature of the MoS$_2$~\cite{brivio2011} at the area under investigation. A diffractogram of the G/MoS$_2$/G sample is presented in \ref{firstimage}(b), showing the two graphene peaks originating from the top and bottom layers, and the peaks of alternating intensity from the MoS$_2$ monolayer.

Once the correct sample area was located, the material was observed in the high-resolution mode, while keeping track of the total accumulated electron dose. All the samples were investigated in similar conditions in terms of vacuum level, dose rate, magnification, and total beam current. Importantly, atomically clean areas were found in the heterostructure areas, which corroborates the earlier observation by Haigh {\em et al.}~\cite{haigh2012}. There interfaces in similar heterostructures were found to be contamination free, and in full contact in cross-sectional images. Similarly to what was observed in our earlier study~\cite{komsa2012}, exclusively sulphur vacancies are created under the electron beam. By observing the rate at which the vacancy concentration increases, one can directly evaluate the total vacancy production cross-section in each case. 

The dramatic effect of the graphene layers in the G/MoS$_2$/G sample can be seen when \ref{firstimage} panels (c) and (d) are compared. In panel (c), the free-standing MoS$_2$ sample is seriously damaged after an electron dose of $2.8\times10^8$ electrons/nm$^2$, and the vacancies have partly rearranged into lines, as described in Ref.~\cite{komsa2013}. Consequently, no information on the atomic structure of the pristine state of the sample can be extracted. With the G/MoS$_2$/G sample (panel (d)) the situation is completely different after the same electron dose. Here, for example, the structure of the MoS$_2$ flake edge can be readily observed at atomic resolution, which would not be possible with the unprotected sample, and the vacancy concentration in the flake remains very low. Panel (e) shows the same HRTEM frame as in panel (d), but after filtering out the graphene contribution and high frequency noise by means of Fourier filtering, which further improves the interpretability of the MoS$_2$ structure by removing the Moir\'{e} effect resulting from the overlay of the MoS$_2$ and the graphene lattices. 

\begin{table}

\caption{{\bf The measured quantities from the four heterostructure samples.} $\Delta N$ is the number of lost S atoms, $N$ the total number of S sites in the investigated area, $\phi$ the accumulated electron dose, and $\sigma$ the vacancy production cross-section as calculated by $\Delta N$/$(N$$\phi )$. The confidence intervals were calculated assuming $N/\sqrt{N}$ errors for the integer quantities and 1\% errors for the electron doses.}

\label{quantities}
\begin{center}
\begin{tabular}{ r|c|c|c|c}
Sample & $\Delta N$ & $N$ & $\phi$ (e/nm$^2$) & $\sigma$ (barn) \\
\hline
  MoS$_2$ & 116 & 3176 & 8.2$\times 10^7$ & 4.5(4)\\
  G/MoS$_2$ & 43 & 2894 & 9.7$\times 10^7$ & 1.5(2)\\
  MoS$_2$/G & 177 & 5294 & 7.0$\times 10^8$ & 0.48(4)\\
  G/MoS$_2$/G & 5 & 3090 & 2.14$\times 10^9$ & 0.008(3) \\
\end{tabular}
\end{center}
\end{table}

For quantifying the damage rate and for studying the contributions of the different damage mechanisms, all the four samples were given an electron dose at which the vacancy concentration increased to up to 4~\%, and HRTEM images before and after the irradiation were compared, taking account of the increase in vacancy concentration. The relevant measured quantities are given in \ref{quantities}, along with the calculated vacancy production cross-sections. The cross-sections are calculated by $\sigma=\Delta N/(N\phi )$, where $\Delta N$ is the number of lost S atoms, $N$ the total number of S sites in the investigated area, and $\phi$ the accumulated electron dose.

\begin{figure}[!h]
	\includegraphics[width=.98\textwidth]{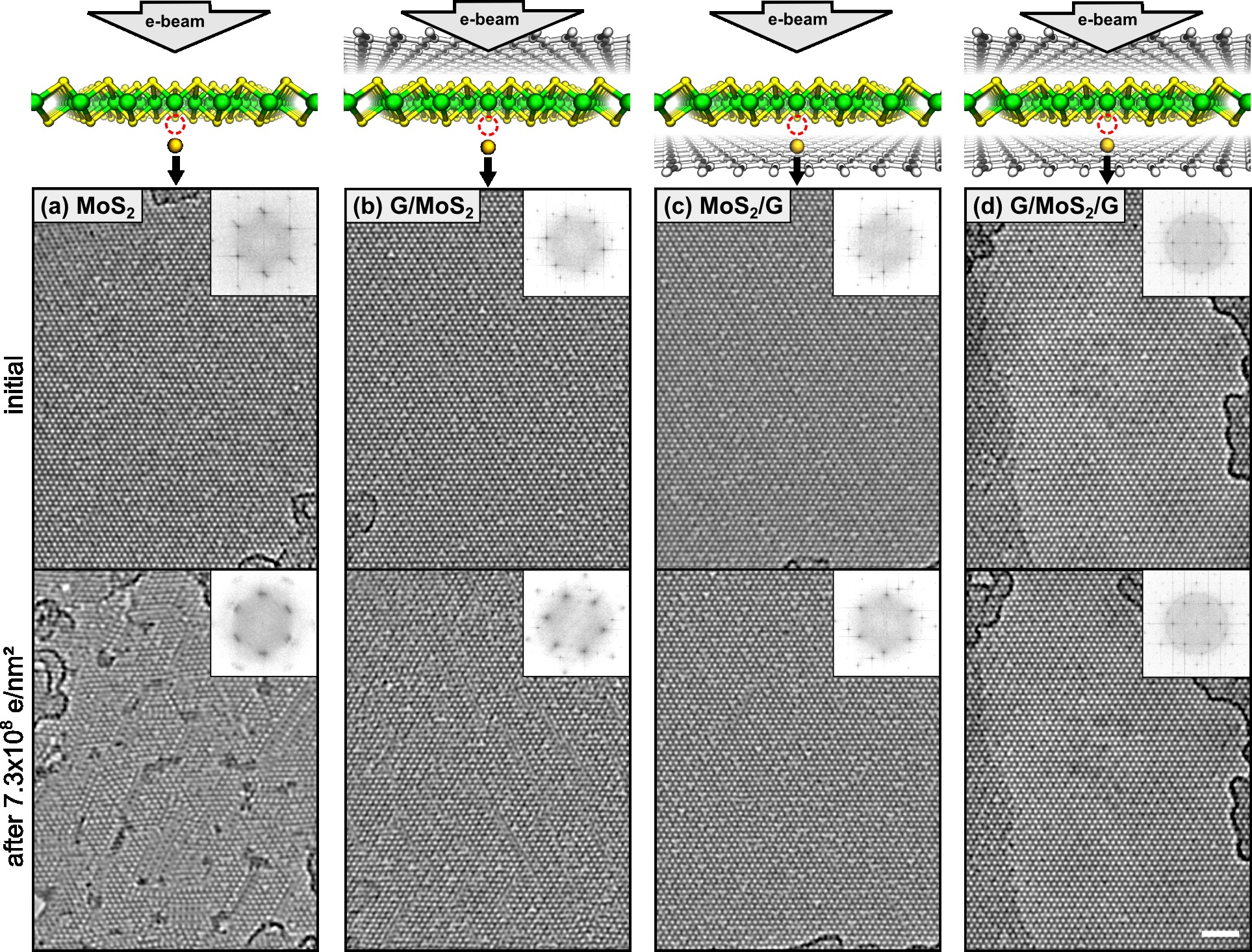}
	\caption{{\bf MoS$_2$/graphene heterostructures exposed to 80~keV electron irradiation in a transmission electron microscope.} The upper HRTEM images show the initial state of the samples, and the lower images the final state after a cumulated electron dose of 7.3 $\times$ $10^8$ e/nm$^2$ at a constant dose rate of 9 $\times$ $10^5$  e/nm$^2$/s. {\bf (a)}, Free-standing MoS$_2$. After the electron dose the sample is severely damaged and the crystalline order is lost, as can be seen from the FFTs (insets of the HRTEM images). {\bf (b)}, Graphene on top (G/MoS$_2$). The damage is reduced from the previous case when the top side is covered with graphene. {\bf (c)}, Graphene on the bottom. Further reduction of damage is observed, as now the graphene layer inhibits sputtering of sulfur atoms. {\bf (d)}, The sandwich (G/MoS$_2$/G). Covering both sides of the MoS$_2$ layer leads to a dramatically reduced damage rate. The scale bar is 2~nm.}
	\label{damageimages}
\end{figure}

The significant differences between all the four sample configurations can be seen in the HRTEM images in \ref{damageimages}, where each case is shown before and after the electron dose of 7$\times 10^8$ e/nm$^2$. The free-standing MoS$_2$ layer is again seriously damaged after this electron dose, resulting in the loss of long-range order, as indicated by the broadening of the lattice maxima in the Fourier transforms. The sample in the G/MoS$_2$ configuration shows reduced damage, but is still far from the pristine state of the material, and the sample in the MoS$_2$/G configuration shows further reduced damage. The G/MoS$_2$/G sample, however, shows dramatically different behavior, and the vacancy concentration remains unchanged after the electron dose which completely destroyed the free-standing target. We would like to point out, that we do see occasional migration steps of vacancies also in the G/MoS$_2$/G configuration (the locations of the vacancies in the G/MoS$_2$/G frames in \ref{damageimages} are not identical). The difference between damage rates in the G/MoS$_2$/G and the G/MoS$_2$ configurations is further demonstrated in Supplementary material Figure~S1 and Video~1, where a spot in the G/MoS$_2$/G sample was located, at which the bottom graphene layer ends, thus allowing the observation and comparison of a fully and half covered parts at the exact same conditions.

The cross-sections for the MoS$_2$, G/MoS$_2$, MoS$_2$/G, and G/MoS$_2$/G configurations were determined to be 4.5(4) barn, 1.5(2) barn, 0.48(4) barn, and 0.008(3) barn, respectively (one barn equals $10^{-10}$~nm$^2$). As compared to the free-standing case, the cross-section is reduced by a factor of 2.9 for the G/MoS$_2$ configuration, 9.3 for the MoS$_2$/G configuration, and 600 for the G/MoS$_2$/G configuration. If one defines the critical electron dose to be the dose at which 5\% of target sulphur atoms are lost, one gets values of 1.12(11)$\times 10^8$ e/nm$^2$, 3.3(5)$\times 10^8$ e/nm$^2$, 1.05(8)$\times 10^9$ e/nm$^2$, and 6(3)$\times 10^{10}$ e/nm$^2$ for the MoS$_2$, G/MoS$_2$, MoS$_2$/G, and G/MoS$_2$/G configurations, respectively. Quantification of damage in the graphene layers is not possible due to the much stronger contrast originating from the MoS$_2$ layer. However, prior studies have shown that graphene is quite robust under 80 keV electron beam, and tends to flexibly reorganize into a closed network in the presence of vacancy type defects~\cite{kotakoski2011,lehtinen2013}.

By comparing the samples in the G/MoS$_2$ and MoS$_2$/G configurations, it can immediately be seen that knock-on damage is not the dominant damage mechanism in MoS$_2$ under the electron beam, contrary to what was assumed in our earlier paper~\cite{komsa2012}. If suppression of knock-on damage is taken to be the only difference between the two configurations, one can estimate the contribution of the knock-on process. A value of 1.1(2)~barn is acquired, which is only 24\% of the total damage cross-section in the free-standing case and consequently 76\% of the damage is of a different origin. This knock-on cross-section is in good agreement with the theoretical prediction of 0.8~barn~\cite{komsa2012}. It should be pointed out that the theoretical value is based on a displacement threshold calculated specifically for sulphurs in the bottom layer, which are easier to displace than the top layer sulphurs.

Comparison of the samples with only single side covered and the free-standing MoS$_2$ sample allows studying the effects of the other damage mechanisms, although clearly isolating the mechanisms is not as straight-forward as in the case of knock-on damage. If one assumes that already covering a single side of the MoS$_2$ layer with graphene suppresses the damage processes caused by electronic excitations and charging, and thus only chemical etching on one surface is active in the MoS$_2$/G configuration, one can further calculate a cross-section of 2.4(6)~barn as the contribution related to electronic excitations and chaging (55\% of the free-standing cross-section). This number, however, should be taken as a rough estimate. In future experiments it would be of great interest to replace the graphene layer on one or both surfaces by a hexagonal boron-nitride layer, which is chemically inert, has comparable mechanical properties to those of graphene, but is an insulating material. Such samples would allow further separation of the different damage mechanisms.

To conclude, we have successfully applied graphene as a coating material for dramatically reducing radiation damage in single layer MoS$_2$ induced by the 80~keV electrons used for high-resolution imaging in AC-HRTEM. Our new sample preparation method allows characterization of the pristine atomic structure of a radiation sensitive material. Moreover, we demonstrated the usefulness of different layered graphene-MoS$_2$ heterostructures in separating different radiation damage mechanisms. The technological implications are obvious: Using graphene as a protective layer is highly advantageous when imaging radiation sensitive materials in a transmission electron microscope, from thin layers to isolated molecules. This expands the applicability of the AC-HRTEM significantly, as radiation damage has been a serious hindrance in fully exploiting the greatly improved resolving power of the aberration-corrected instrument. The total vacancy creation cross-section was reduced 600 fold going from a free-standing MoS$_2$ layer (4.5(4) barn) to the sandwiched configuration (0.008(3) barn). The contribution of knock-on damage was separated from the total damage cross-section (1.1(2) barn), and was observed to only partially explain the total accumulated damage. An estimate on the cross-section related to electronic excitations was deduced (2.4(6) barn), although heterostructures of different materials would be useful in addressing this effect more reliably. Other types of displacing radiation were not explicitly studied here, but our results suggest that graphene can be used also more generally in protecting surfaces from radiation damage.

\begin{acknowledgments}
The authors acknowledge the financial support by the DFG (German Research Foundation) and the Ministry of Science, Research and the Arts (MWK) of Baden-Wuerttemberg in the frame of the SALVE (Sub Angstrom Low-Voltage Electron microscopy) project. O.L. acknowledges financial support from the Finnish Cultural Foundation.

{\bf Author contributions}
G.A.S., S.K., and U.K conceived and planned the experiments. M.S. prepared the samples under the supervision of G.A.S. and S.K. S.K. executed the HRTEM imaging under the supervision of U.K. All the authors contributed to the analysis of the data. O.L. wrote the manuscript with assistance from U.K. and all the authors.
\end{acknowledgments}

\end{document}